\documentclass[journal=jacs,manuscript=article]{achemso}

\usepackage[version=3]{mhchem} 
\usepackage[T1]{fontenc}       

\newcommand{\onlinecite}[1]{\hspace{-1 ex} \nocite{#1}\citenum{#1}} 
\usepackage{chemfig}

\setdoublesep{0.35700 em}  
\setatomsep{1.78500 em}    
\setbondoffset{0.18265 em} 
\newcommand{\bondwidth}{0.06642 em} 
\setbondstyle{line width = \bondwidth}

\author{Carolin K\"onig}
\affiliation[KTH]
{Department of Theoretical Chemistry and Biology, KTH Royal Institute of Technology, Roslagstullsbacken 15, SE-116 91 Stockholm, Sweden}
\author{Robin Sk\aa nberg}
\affiliation[VIZ]
{Link\"opings University, Department for Media and Information Technology, Norrk\"oping, Sweden}
\author{Ingrid Hotz}
\affiliation[VIZ]
{Link\"opings University, Department for Media and Information Technology, Norrk\"oping, Sweden}
\author{Anders Ynnerman}
\affiliation[VIZ]
{Link\"opings University, Department for Media and Information Technology, Norrk\"oping, Sweden}
\author{Patrick Norman}
\affiliation[KTH]
{Department of Theoretical Chemistry and Biology, KTH Royal Institute of Technology, Roslagstullsbacken 15, SE-116 91 Stockholm, Sweden}
\author{Mathieu Linares}
\email{linares@kth.se}
\affiliation[KTH]
{Department of Theoretical Chemistry and Biology, KTH Royal Institute of Technology, Roslagstullsbacken 15, SE-116 91 Stockholm, Sweden}
\altaffiliation{Swedish e-Science Research Centre (SeRC), KTH Royal Institute of Technology, 104 50 Stockholm, Sweden}

\title[Amyloid Binding Sites]{Binding Sites for Luminescent Amyloid Biomarkers from non-Biased Molecular Dynamics Simulations}

\abbreviations{}

\begin{document}

%
%


\begin{abstract}
A very stable binding site for the interaction between an pentameric oligothiophene and an amyloid-$\beta$(1--42) fibril has been identified by means of \textit{non-biased} molecular dynamics simulations. In this site, the probe is locked in an all-\textit{trans} conformation with a Coulombic binding energy of 1,200 kJ/mol due to the interactions between the anionic carboxyl groups of the probe and the cationic $\varepsilon$-amino groups in the lysine side chain. Upon binding, the conformationally restricted probes show a pronounced increase in molecular planarity. This is in-line with the observed changes in luminescence properties that serve as the foundation for their use as biomarkers.
\end{abstract}

Amyloid fibrils are pathological hallmarks of a number of human neurodegenerative diseases such as Alzheimer's and Parkinson's.\cite{Ross2004,Chiti2006,Radford2012} These fibrils are shown to interact with luminescent conjugated oligothiopene (LCO) biomarkers, enabling an early-stage, specific, detection of the associated diseases\cite{Nilsson2007, Aslund2009a, Simon2014, Klingstedt2015a, Shirani2015} as the LCOs have proven to distinguish between separate morphotypes of protein aggregation that are related to different heterogeneous phenotypes.\cite{Simon2014} 

To achieve rational biomarker design, however, it is indispensable to have a detailed microscopic knowledge of the binding between the biomarker and the amyloid fibril. Although in principle well suited for theoretical studies, simulations of the binding site have been hampered by a lack of knowledge of fibrillar structures. Recent advances in the structural biology have changed this situation,\cite{Gremer2017} allowing us to perform large-scale \textit{non-biased} molecular dynamics simulations of the binding of LCOs to amyloid fibrils. Adopting one of the prototypical anionic LCOs, our simulations reveal a binding site that convincingly reproduces all currently known experimental observations and we believe that it is representative for this class of probes. 

The recent experimental fibrillar structure studies all give evidence for a double-stack of amyloids that is tilted\cite{Walti2016, Colvin2016, Tycko2016, Gremer2017} but differ in their predictions of the relative stack positions. Refs.~\onlinecite{Walti2016, Colvin2016} are based on nuclear magnetic resonance (NMR) measurements and find a flexible and non-structured terminus. In contrast, the structure obtained from cryo-electron microscopy (cryo-EM) reported in Ref.~\onlinecite{Gremer2017} shows a very ordered organization of the entire protein. This study furthermore provides a direction for the tilt and suggests an approximate $2_1$ screw symmetry. 
 
\begin{figure}[ht]
\small
\
\chemfig{[:-60]^\ominus O-C([:215]=O)-[:-18]
*5(=-=
(-*5(-S-
(-*5(=-=
(-*5(-S-
(-*5(=-=
(- C ([-15]=O)-[:-45]O^\ominus)
-S-))
=
-
(--[:90]C([:120]=O)-[:20]O^\ominus)
=))
-S-))
=
(--[:90]C([:20]=O)-[:120]O^\ominus)
-=)
)
-S-)
}
\includegraphics[width=0.65\textwidth]{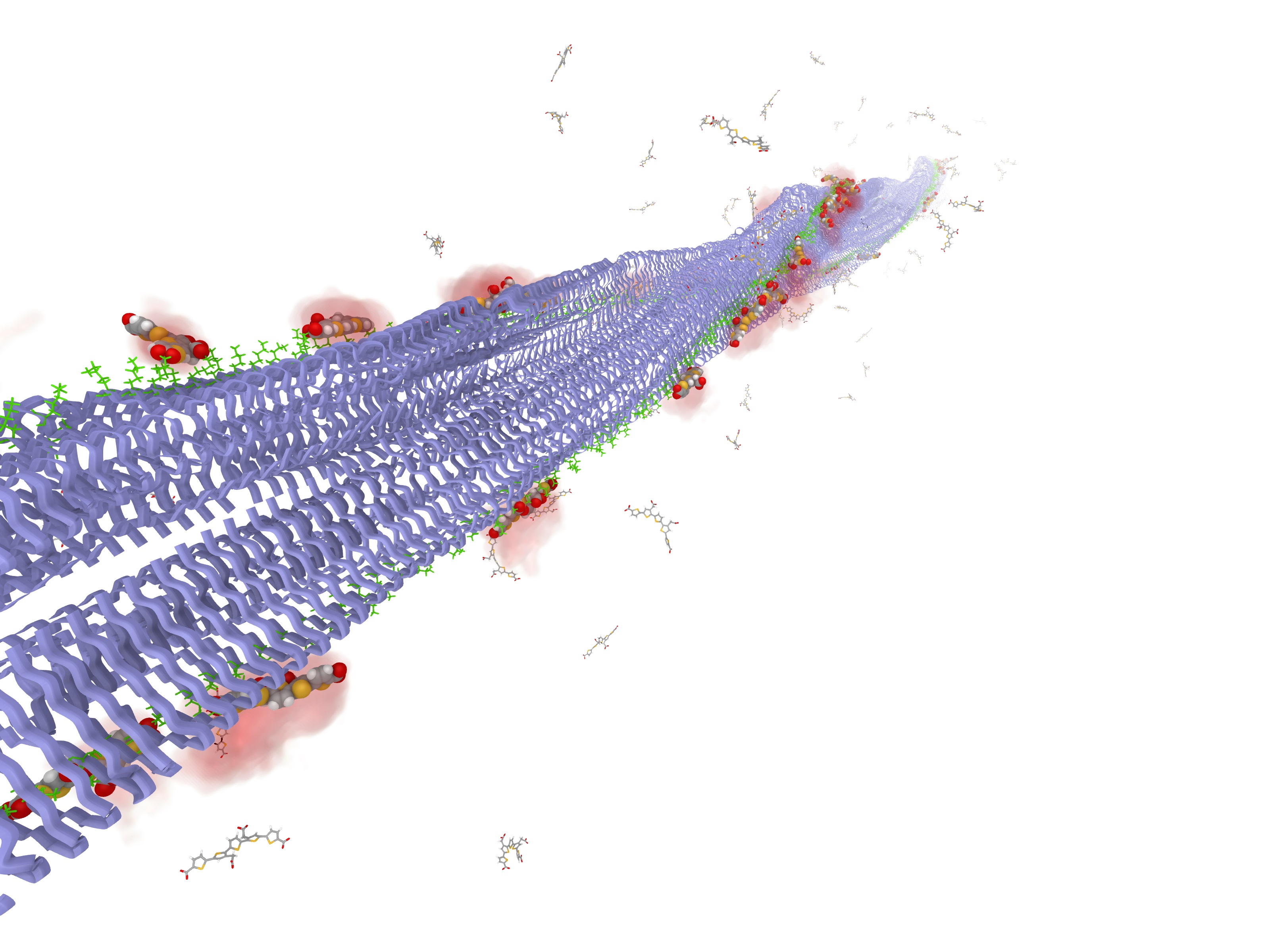}
\caption{
Top: Structure illustration of adopted LCO probe named p-FTAA.
Bottom:
Simulation snapshot after 40~ns. The protein and lysine 16 groups are depicted in blue and green, respectively. Probes showing large short-range Coulombic interactions with the fibril are highlighted. The red shaded areas show density volumes for probes with Coulombic binding energies and planarities exceeding 500 kJ/mol and 2.5, respectively. 
}
\label{fig:lco}
\end{figure}

Earlier docking studies include (i) the work by Sch{\"u}tz \textit{et al.} investigating the interactions between Congo red,\cite{Schutz2011} a more standard amyloid biomarker, and amyloid fibrils formed by the prion domain of the fungal HET-s protein as well as (ii) the work by Herrmann \textit{et al. } investigating the interactions between the LCO adopted here\cite{Herrmann2015} (see molecular structure in Figure~\ref{fig:lco}) and a misfolded and aggregated form of the cellular prion protein (PrP$^\mathrm{C}$). Both these studies showed significant interactions between the negatively charged sulfonate and carboxyl moieties of the probes, respectively, and the positively charged lysine residues of the respective protein models. In contrast and comparison to these earlier studies, we apply a \textit{non-biased} and fully theoretical approach using the recently discovered tilted structure of the amyloid fibril.\cite{Gremer2017}

We have constructed the structural model from one chain of the PDB entry 5OQV.\cite{Gremer2017} We duplicated this chain multiple times and applied the approximate $2_1$ screw symmetry reported in Ref.~\onlinecite{Gremer2017}. With 253 chains we obtained a one-half helix turn, allowing us to apply periodic boundary conditions. We optimized the chain--chain distance with help of a semi-isotropic barostat to accommodate the room temperature expansion as compared to the ultra-cold conditions in which the structure was obtained.\cite{Gremer2017} We thereby obtained a simulation box of ca.\ 63~nm along the axis of the fibril, corresponding to a pitch of 126~nm for the amyloid fibril helix. The model proved stable in our molecular dynamics simulations lasting up to 9.8~ns. Based on this structure, we generated a simulation box containing the fibril and 61 tracked LCO probes in a similar fashion. After $NPT$-equilibration, we performed a $NVT$-simulation for 40~ns on the full system. The computational details for the molecular dynamics simulations with Gromacs\cite{Berendsen1995, Lindahl2001, vanderSpoel2005} as well as the applied force fields\cite{Brooks1983, Brooks2009, Bjelkmar2010, Jorgensen1983, Sjoqvist2014} and more details concerning the model setup and supporting  calculations\cite{Frisch2016, Yanai2004, Petersson1988, Petersson1991} can be found in the supplementary information.

After only a few nanoseconds of simulation time, some of the chromophores  started to bind to the amyloid fibril. 
All observed binding processes are initially governed by electrostatic interactions of the negatively charged carboxyl groups of the probe with positively charged side chains on the surface of the amyloid fibril. This is in a vast majority of the cases the lysine 16 group.  
Once, the connection is established, the probe stays very local to its  original interaction site. At later stages, we observe that the probes lie flat down on the protein surface, which gives rise to additional hydrophobic (Lennard-Jones) binding energy between the probe and the amyloid fibril. An example of this binding behavior is illustrated in Figure S-3 in the supplementary material.

After 40~ns, we found 25 LCO probes with significant short-range Coulomb interaction with the fibril and 22 of these were located at the lysine 16 interaction side, as highlighted in Figure~\ref{fig:lco} with the VAMD visualization environment (see details in supplementary information). The other three LCO molecules binded via electrostatic interactions with the arginine 5 groups. These interactions are, however, weaker than those with  the lysine 16 side chains.

\begin{figure}[ht]
\includegraphics[width=\textwidth]{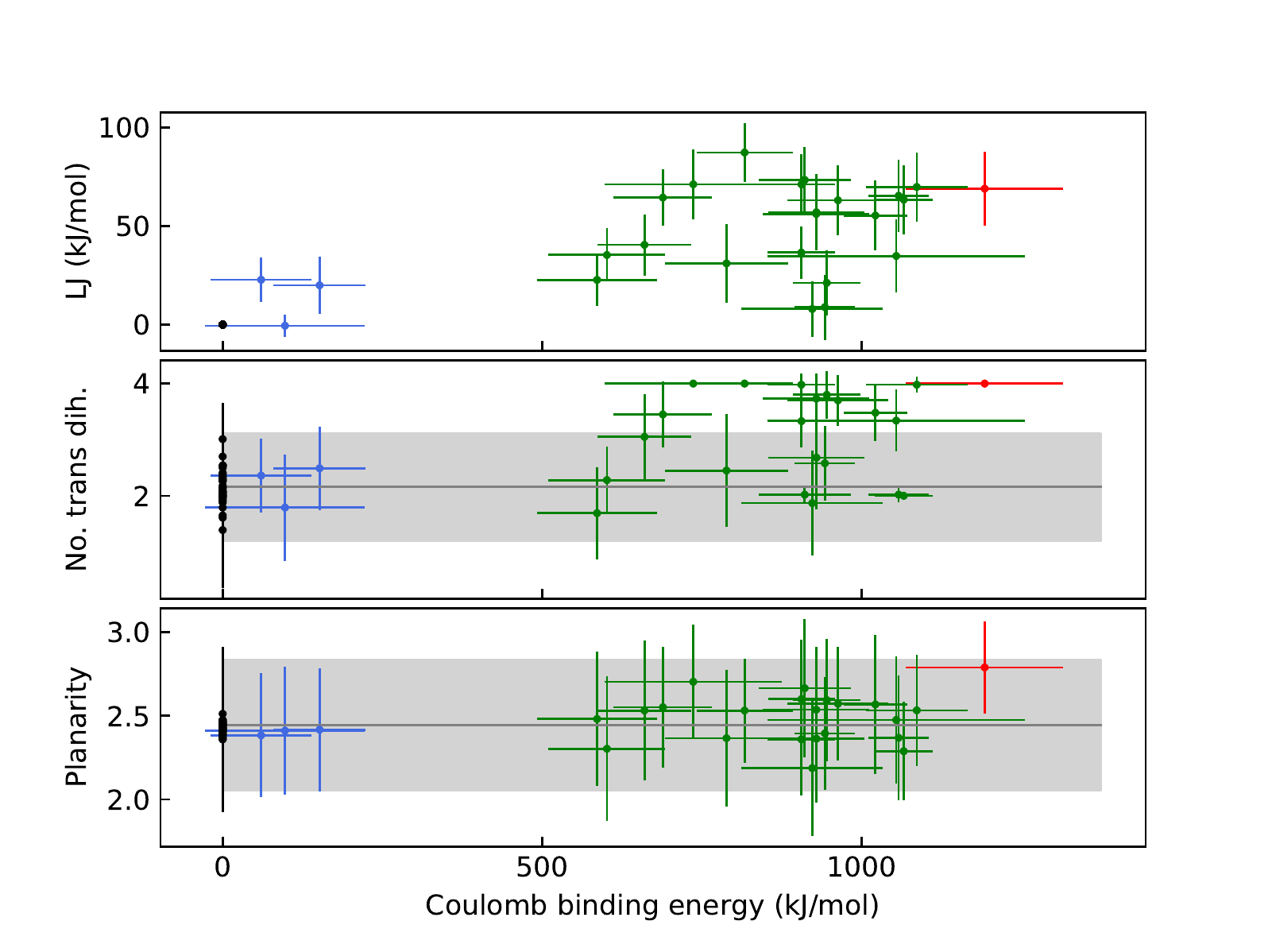}
\caption{Correlation of Coulomb binding energy with (i) the Lennard-Jones binding energy (LJ, top), (ii) number of trans dihedral angles (middle), and (iii) planarity (bottom) for all 61 tracked probes. Circles show averages over simulation time region 39--40~ns and bars represent RMS deviations. The grey areas mark reference data for a single probe in water. 
\label{fig:scatter}}
\end{figure}

To quantify the overall behavior of the probes over time, we studied the system over the last nanosecond of the simulation, i.e., from 39--40~ns in real time. In Figure~\ref{fig:scatter}, we present the correlation of the different properties of interest with the Coulomb binding energy for all monitored probes separately and within this time-window. Time-averaged quantities are shown with filled circles and bars mark the corresponding root-mean-square (RMS) deviations.

The upper panel of Figure~\ref{fig:scatter} shows the correlation between the Coulomb binding energy between the individual probes and the protein and the corresponding Lennard-Jones binding energy. We can clearly identify three different categories of probe molecules: (i) Colored in green and red, the 22 probe molecules bound to the lysine 16 group with average Coulomb binding energies in the range of 500--1200 kJ/mol and Lennard-Jones binding energies ranging up to $80$ kJ/mol; (ii) Colored in blue, the three probes interacting with the arginine 5 group, showing Coulomb binding energies of 50--300 kJ/mol and Lennard-Jones binding energies ranging up to $20$ kJ/mol; (iii) Colored in black, the unbound probes that have no interaction energy energies, are collectively represented by a single black dot. We note that probes with strong Coulombic binding also have sizable Lennard-Jones binding energies, but that there is no clear correlation between the two energy contributions for probes within any of the three categories. 

The mid panel of Figure~\ref{fig:scatter} shows the correlation between the Coulomb binding energy and the number of \textit{trans}-dihedral S--C--C--S angles in the probe. We observe a pronounced increase in the number of \textit{trans}-dihedral angles for the category of probes bound to lysine 16.  

\begin{table}[ht]
\caption{Time-averaged planarity and number of \textit{trans}-dihedral angles together with the respective RMS deviations. Results are presented for the different categories of probes over the time range from 39 to 40~ns. Converged results for a single probe in aqueous solution 
are provided as well as results for the strongly bound p-FTAA probe for the large and small model system. 
\label{tab:stats}}
\begin{tabular*}{\textwidth}{@{\extracolsep{\fill}}lrrrrr}
\hline
\hline
                   & no.~mol   & time (ns) & & planarity  & no.~trans \\
\hline                                 
Single p-FTAA in water  &   1     &  100    &              & 2.45$\pm$0.39  & 2.16$\pm$0.94 \\
Bound LCOs to LYS16     &  22     &  1  & & 2.49$\pm$0.40  & 3.07$\pm$0.97 \\
Bound LCOs to ARG5      &   3     &  1  & & 2.40$\pm$0.37  & 2.21$\pm$0.85 \\
Unbound LCOs           &  36     &  1  & & 2.43$\pm$0.38  & 2.14$\pm$0.95 \\
\hline
Bound p-FTAA in large model & 1 & 1 & & 2.79$\pm$0.28 & 4.00$\pm$0.00 \\
Bound p-FTAA in large model & 1 & 23.4 & & 2.83$\pm$0.29 & 4.00$\pm$0.06 \\
Bound p-FTAA in small model & 1 & 62.4 &                        & 2.84$\pm$0.29 & 4.00$\pm$0.07 \\
\hline
\hline
\end{tabular*}
\end{table}

The lower panel of Figure~\ref{fig:scatter} shows the correlation of probe planarity, as defined in Ref.~\onlinecite{Sjoqvist2014a}, with respect to the Coulomb binding energy. For a given molecular probe configuration, this planarity measure results in a value between 0 and 4, corresponding to having all four dihedral angles equal to 90$^\circ$ and 0$^\circ$ or 180$^\circ$, respectively. As reference, the solid horizontal gray line and shaded light gray area represents that averaged planarity and its RMS deviation for a 100 ns single-probe trajectory in aqueous solution---their values are listed in Table~\ref{tab:stats} as equal to 2.45 and 0.39, respectively. From the data in the table, one can conclude that there is no statistical significant changes in planarity to be found for any of the three probe categories in comparison with the reference results for an isolated probe in aqueous solution. This stands in contrast with the experimental observations of red-shifted excitation spectra associated with increasing planarity,\cite{Sjoqvist2014a} so clearly none of the ensemble categories of probes can provide a realistic representation of the probe-to-fibril binding.

\begin{figure}[ht]
\includegraphics[height=0.25\textheight]{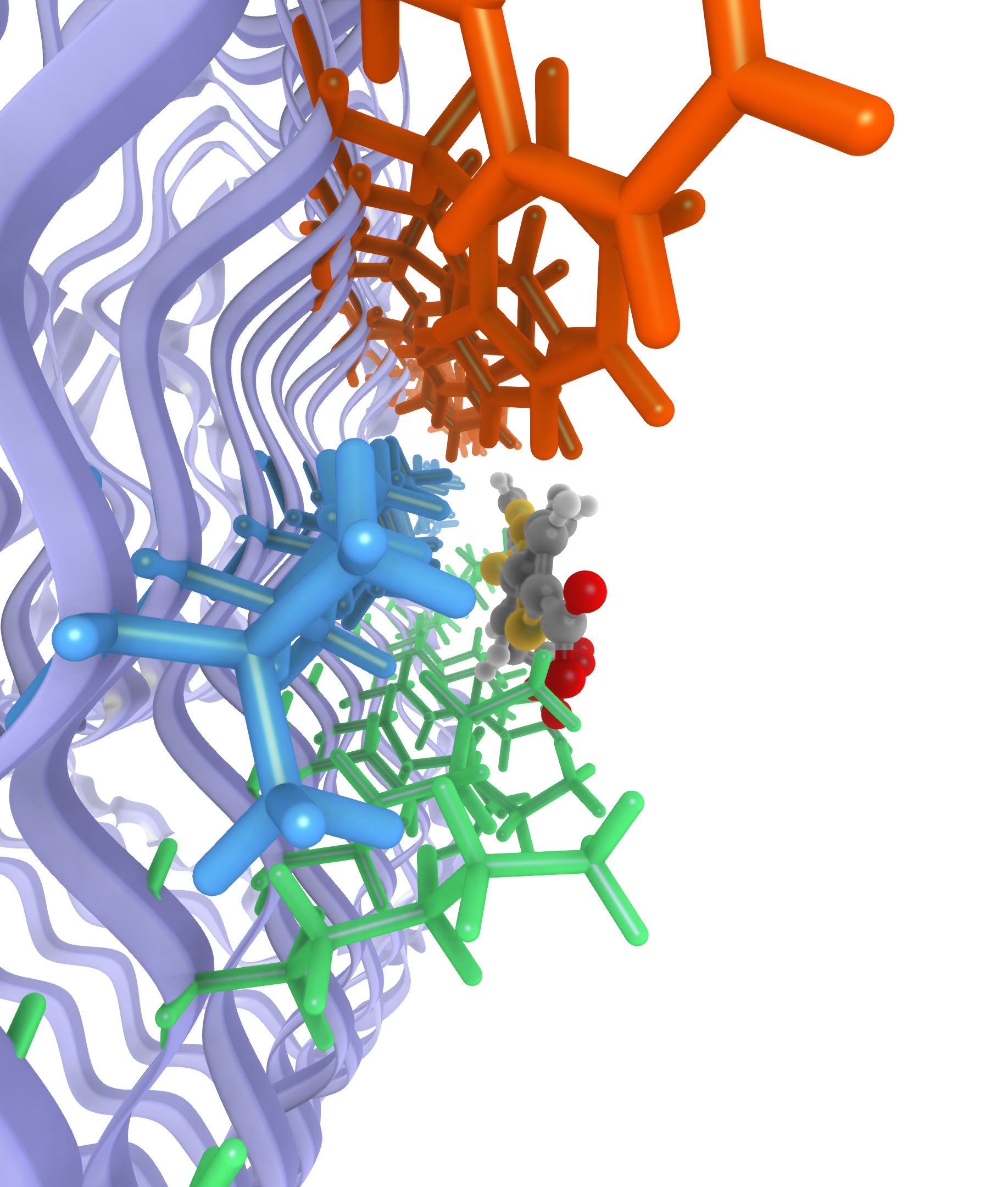}
\includegraphics[height=0.25\textheight]{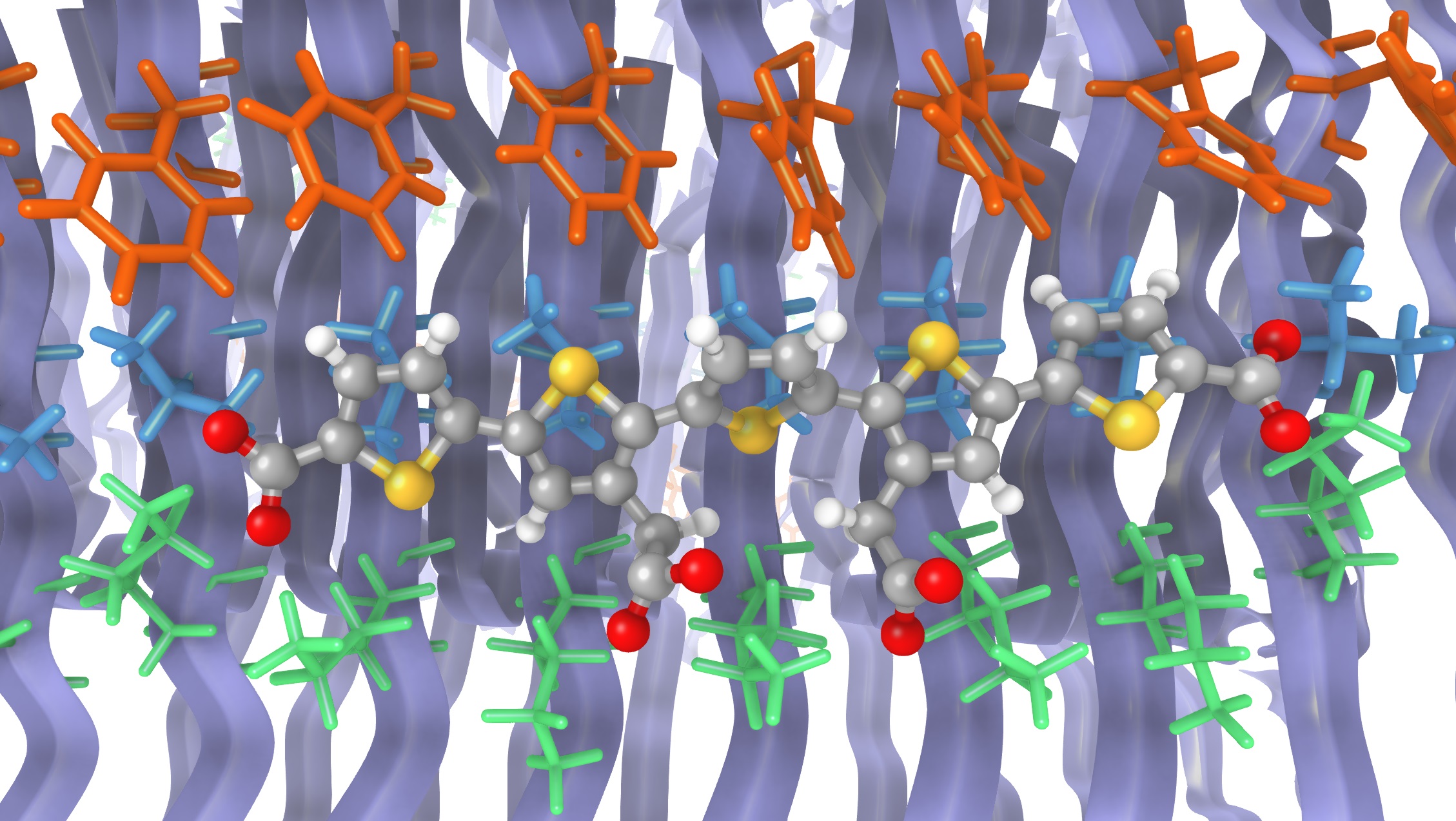}
\caption{Strongest binding motif from different viewpoints. Positively charged lysine 16 groups are shown in green and the hydrophobic side chains of valine 18 and phenylanaline 20 in blue and orange respectively. 
\label{fig:binding-site}}
\end{figure}

However, within the category of lysine 16 bound probes, one LCO stands out, namely the one highlighted in red in Figure~\ref{fig:scatter} and henceforth referred to as p-FTAA$^{*}$. This binding conformation is depicted in Figure~\ref{fig:binding-site} and has a Coulomb binding energy as high as 1200$\pm$100 kJ/mol combined with a short-range Lennard-Jones binding energy of 70$\pm$20 kJ/mol. As presented in Figure~\ref{fig:binding-site}, this binding site contains in addition to the positive lysine 16 groups interacting with the carboxyl groups of the LCO also two hydrophobic side-chains, namely valine 18 and phenylalanine 20---the side-chains for the residues 17 and 19 are pointing inwards the fibril and are therefore not shown. 

Compared to the other lysine-bonded probes, p-FTAA$^{*}$ exhibits higher degrees of planarity and \textit{trans}-conformation as seen in Table~\ref{tab:stats}. With reference to aqueous solution, the planarity of p-FTAA$^{*}$ is increased by 0.3--0.4. The binding conformation is very stable and once formed it was never observed to disintegrate during the full simulation time. To collect further evidence for the stability of this binding site, we performed longer molecular dynamics simulations on a smaller model system (see Figure S-2 in the supplementary material). Using this reduced-size system, we could afford a simulation time exceeding 60~ns, during which the p-FTAA$^{*}$ conformation remained stable. The statistics from this simulation are shown at the bottom row of Table~\ref{tab:stats} and they are seen to be in perfect agreement with those obtained in the shorter simulation of the full system (table row above). Hence, we consider these molecular dynamics simulations converged and it is with some confidence that we present this as the main binding site for anionic LCO probes interacting with amyloid fibrils.

The proposed binding site is in agreement with experimental observations on multiple accounts:
\begin{itemize}
\item The UV-vis absorption spectrum of the bound LCO probe is shifted by about 30~nm (0.19~eV).\cite{Shirani2015} Previous work\cite{Sjoqvist2014a} reports on a correlation between the planarity and the absorption maximum of LCO dyes in water. For p-FTAA, a linear fit with a slope of 53.5~nm is obtained (see Figure~S-4 in the supplementary material), which, with an increased planarity for p-FTAA$^{*}$ of about 0.3--0.4, corresponds to a red-shift of about 15--20 nm. This rough estimate points into the right direction, leaving out important aspects in the spectrum calculation such as effects of vibrations, conformational restriction in the binding site, and electronic polarization due to the environment. Such an extensive spectrum study lies beyond the scope of the present communication.
\item It is known that the spacing between the carboxyl groups is essential for spectral discrimination of different fibrils.\cite{Klingstedt2015a} This finding suggests a regular and specific interaction of the carboxyl groups with the fibril, just as we observe in every single binding event in this study.
\item The interactions between negatively charged probe groups and lysine have been proposed before as main interaction sites for both Congo-red\cite{Schutz2011} and LCO probes\cite{Herrmann2015, Schutz2017} in combination with prions. These results were obtained from evidence-based docking, based on more approximate structures than what we employ in the present study.
\end{itemize}

In summary, we propose a site for the binding of anionic luminescence biomarkers with amyloid-$\beta$(1--42) fibrils based on \textit{non-biased} molecular dynamics simulations. The found binding site is very stable, leading to conformational restrictions of the probe that are in qualitative agreement with observed spectroscopic shifts upon the aggregation of LCO probes with a recombinant A$\beta$(1--42) peptide.\cite{Shirani2015} Convinced that we have found a realistic binding site for this class of probes, we believe that this elucidation of the probe--protein interactions on atomistic level will be crucial for the rational design of future and improved biomarkers for amyloid fibrils. 
 
\begin{acknowledgement}
CK acknowledges funding by a Marie Sk{\l}odoswka--Curie International Fellowship ``FreezeAlz'' by the European Commission (Grant No.\ 745906). ML, RS, IH and AY thanks SeRC (Swedish e-Science Research Center) for funding. PN thanks the Swedish Research Council (Grant No. 621-2014-4646) for funding. The  Swedish National Infrastructure for Computing (SNIC) at National Supercomputer Centre (NSC) and Center for High Performance Computing (PDC) are acknowledged for providing computer resources.
\end{acknowledgement}

%


\providecommand{\latin}[1]{#1}
\makeatletter
\providecommand{\doi}
  {\begingroup\let\do\@makeother\dospecials
  \catcode`\{=1 \catcode`\}=2 \doi@aux}
\providecommand{\doi@aux}[1]{\endgroup\texttt{#1}}
\makeatother
\providecommand*\mcitethebibliography{\thebibliography}
\csname @ifundefined\endcsname{endmcitethebibliography}
  {\let\endmcitethebibliography\endthebibliography}{}

\end{document}